\documentclass[prb,twocolumn,showkeys,showpacs,preprintnumbers,amsmath,amssymb,floatfix]{revtex4}

\usepackage{graphicx}
\usepackage{dcolumn}
\usepackage{bm}
\usepackage{dsfont}

\begin{document}

\title{Strain, Magnetic Anisotropy, and Anisotropic Magnetoresistance in (Ga,Mn)As on High-Index Substrates:
Application to (113)A-Oriented Layers}

\author{L. Dreher\footnote{dreher@wsi.tum.de, present address: Walter Schottky Institut, Technische Universit\"at M\"unchen, 85748 Garching, Germany}}
\author{D. Donhauser}
\author{J. Daeubler}
\author{M. Glunk}
\author{C. Rapp}
\author{W. Schoch}
\author{R. Sauer}
\author{W. Limmer}
\affiliation{Institut f\"ur Halbleiterphysik, Universit\"at Ulm, 89069 Ulm,
Germany}


\date{\today}

\begin{abstract}
Based on a detailed theoretical examination of the lattice distortion in high-index epilayers in terms of
continuum mechanics, expressions are deduced that allow the calculation and experimental determination of the
strain tensor for $(hhl)$-oriented (Ga,Mn)As layers. Analytical expressions are derived for the strain-dependent
free-energy density and for the resistivity tensor for monoclinic and orthorhombic crystal symmetry,
phenomenologically describing the magnetic anisotropy and anisotropic magnetoresistance by appropriate
anisotropy and resistivity parameters, respectively. Applying the results to (113)A orientation with monoclinic
crystal symmetry, the expressions are used to determine the strain tensor and the shear angle of a series of
(113)A-oriented (Ga,Mn)As layers by high-resolution x-ray diffraction and to probe the magnetic anisotropy and anisotropic magnetoresistance at 4.2 K by
means of angle-dependent magnetotransport. Whereas the transverse resistivity parameters are nearly unaffected
by the magnetic field, the parameters describing the longitudinal resistivity are strongly field dependent.
\end{abstract}

\pacs{75.50.Pp, 75.47.-m, 75.30.Gw}

\keywords{(Ga,Mn)As; high-index substrates; magnetic anisotropy; strain; magnetotransport}

\maketitle

\section{Introduction}
For the past two decades, dilute magnetic semiconductors, and in particular ferromagnetic semiconductors,
have been attracting considerable attention due to their exceptional physical properties as well as their
potential applicability in information technology. Ferromagnetism mediated by delocalized p-type
carriers\cite{PRB63_195205} could be implemented in the standard semiconductor GaAs by incorporating magnetic
Mn atoms on Ga sites, cf. Ref.~\onlinecite{RMP78_809} and references therein. Since the highest Curie
temperature reported so far is 185~K,\cite{PRL101_077201} the application of (Ga,Mn)As in electronic devices
operating at room temperature seems to be doubtful. Nevertheless, due to its specific electronic and magnetic
properies, (Ga,Mn)As represents an ideal test system for future spintronic applications.

Magnetic anisotropy (MA) and anisotropic magnetoresistance (AMR) are well established key features of (Ga,Mn)As,
largely arising from spin-orbit coupling in the valence band. Consequently, both, MA and AMR strongly depend
on crystal symmetry and strain. On the one hand, this dependence defines the magnetic hard and easy
axes,\cite{RMP78_809,PRB79_195206,PRB78_045203} controlling the orientation of the magnetization and thus the
electrical resistivity. On the other hand, it offers a method to intentionally manipulate the magnetic properties,
e.g. by applying external strain via piezoelectric actuators.\cite{PRB78_045203} In any case, a quantitative
description of the relation between crystal structure and MA and AMR, ideally by means of analytical expressions,
is imperative.

So far, most of the published work focuses on (Ga,Mn)As grown on (001)-oriented substrates while only few publications
report on (Ga,Mn)As grown on high-index substrates such as (113) and
(114),\cite{PE10_206,PRB72_115207,APL89_012507,PRB74_205205,JAP103_093710} where the description of the MA and
the AMR is more complicated due to the reduced crystal symmetry. Potential applications of high-index (Ga,Mn)As
are ridge structures with (113)A sidewalls and (001) top layers,\cite{MJ37_1535} which could be used, e.g. for memory devices exploiting different coercitive fields of the sidewalls and top layers. Yet,
a coherent theoretical description of the MA and AMR for layers on high-index substrates, which
takes into account the symmetry of the strain tensor $\bar{\varepsilon}$, is still missing.

In this work, we present a concise phenomenological description of the MA and AMR for $(hhl)$-oriented
ferromagnetic layers. To this end, we describe the structural properties of epitaxial high-index
layers in terms of continuum mechanics and apply the theoretical results to the case of (113)A orientation.
High-resolution x-ray diffraction (HRXRD) measurements were performed on a series of (113)A-oriented (Ga,Mn)As
layers to quantitatively determine the epitaxial strain and symmetry of the layers. The MA and AMR were
investigated by angle-dependent magnetotransport measurements. In agreement with previous studies on
(113)A-oriented (Ga,Mn)As,\cite{APL89_012507,PRB74_205205,JAP103_093710} we observe a uniaxial MA
along the [001] direction, which can now be explained in the light of our theoretical model. Whereas the transverse
resistivity parameters turn out to be nearly constant, a systematic dependence of the longitudinal
resistivity parameters on the strength of the external magnetic field is found.

The paper is organized as follows: In Sec.~\ref{sec:structure}, we calculate the strain tensor
$\bar{\varepsilon}$ for $(hhl)$-oriented layers and provide formulae that allow experimental access to
$\bar{\varepsilon}$. The results are applied to (113)-oriented (Ga,Mn)As samples where the Bravais lattice
is base-centered monoclinic. An extension of the theory to partially relaxed layers is given in
Appendix~\ref{Appendix_partially_relaxed}. In Sec.~\ref{sec:MA}, we present a phenomenological expression for
the MA taking into account the specific form of the strain tensor for $(hhl)$-oriented layers. In
Sec.~\ref{sec:AMR}, we derive expressions for the longitudinal and transverse resistivities $\rho_\mathrm{long}$
and $\rho_\mathrm{trans}$, respectively, which apply to monoclinic crystal symmetry and current direction
along $[33\bar{2}]$. The resistivity tensors for monoclinic and orthorhombic symmetry can be found in
Appendix~\ref{Appendix_resistivity_tensors}. They are required in the derivation of $\rho_\mathrm{long}$ and
$\rho_\mathrm{trans}$ for arbitrary current directions. In Sec.~\ref{subsec:HRXRD_exp} and
Sec.~\ref{subsec:Exp}, the experimental results of the HRXRD and magnetotransport studies are presented.

\section{Theoretical Considerations\label{sec:theory}}
Crystal symmetry and epitaxial strain strongly influence the MA and the AMR. Therefore, we start with a detailed theoretical examination of the lattice distortion of $(hhl)$ layers in terms of continuum mechanics and apply the general results to (113)-oriented layers (Sec.~\ref{sec:structure}). Based on these results, we present phenomenological expressions for the MA in strained high-index ferromagnetic layers (Sec.~\ref{sec:MA}). We present the resistivity tensors for monoclinic and orthorhombic symmetry (Appendix~	\ref{Appendix_resistivity_tensors}) and calculate $\rho_\mathrm{long}$ and $\rho_\mathrm{trans}$ for monoclinic symmetry and current direction along $[33\bar{2}]$ (Sec.~\ref{sec:AMR}).
\subsection{Structural Properties\label{sec:structure}}

\subsubsection{Distortion of Arbitrarily Oriented Epitaxial Layers\label{subsec:distortion}}

The strain in an epitaxial layer can unambiguously be described by the distortion tensor $\bar{A}$ with components
$A_{ij}=\mathrm{d}u_i /\mathrm{d}x_j$, where $\boldsymbol{u}$ is the mechanical displacement field and $x_i$ denote
Cartesian coordinates along the cubic axes. In order to calculate the distortion of epitaxial layers grown on
arbitrarily oriented cubic substrates, we decompose $\bar{A}$ according to Hornstra and Bartels \cite{JCG44_513}
\begin{equation}
\bar{A}=\mathds{1}\varepsilon_\mathrm{h}+ \boldsymbol{a}\otimes\boldsymbol{n},
\label{eq:A_vec_a}
\end{equation}
where $\otimes$ denotes the dyadic product and
\begin{equation}
\varepsilon_\mathrm{h}=\frac{a_\mathrm{s}-a_\mathrm{l}}{a_\mathrm{l}}
\label{eq:epsilon_h}
\end{equation}
is the isotropic strain that compresses (expands) the cubic unit cell of the layer to the size of the substrate's
unit cell; $a_\mathrm{s}$ and $a_\mathrm{l}$ denote the lattice parameters of substrate and relaxed layer, respectively.
$\boldsymbol{n}$ is the unit vector perpendicular to the interface and  $\boldsymbol{a}$ is a vector that represents
the anisotropic distortion of the layer.
\begin{figure}[htp]
\includegraphics[]{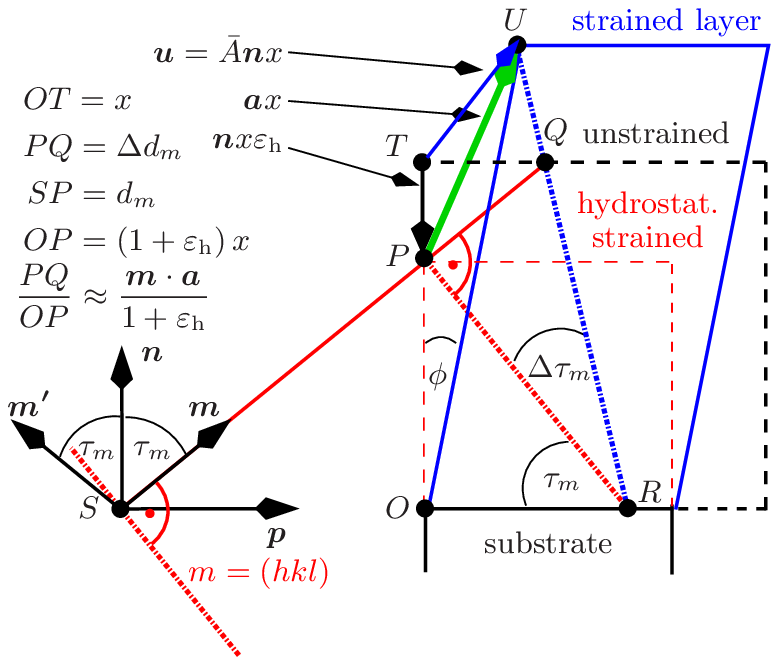}
\caption{(color online) The points $S$, $O$, and $R$ are located within the interface between substrate and layer.
$\boldsymbol{n}$, $\boldsymbol{p}$, $\boldsymbol{m}$, and $\boldsymbol{m'}$ are unit vectors. $\boldsymbol{n}$ and
$\boldsymbol{p}$  are normal and parallel to the interface, respectively; $\boldsymbol{m}$ and $\boldsymbol{m'}$ are
perpendicular to the lattice plane $m=(hkl)$ and $m'=(h'k'l')$. They lie within the plane spanned by $\boldsymbol{n}$
and $\boldsymbol{p}$ and are symmetrically oriented with respect to $\boldsymbol{n}$ ($\tau_m=\tau_{m'}$).\\
Before any strain is being applied, the point $T$ has the distance $x$ from the interface. The displacement
to its final position $U$ can be described as a superposition of a hydrostatic compression of the (unstrained) layer
($T\rightarrow P$) and a displacement by $x\boldsymbol{a}$ ($P\rightarrow U$), cf. Eq.~\eqref{eq:A_vec_a}. The hydrostatically
compressed layer has the same lattice plane spacing as the substrate and therefore Eq.~\eqref{eq:Dd/d} and \eqref{eq:Dtau}
can be derived considering only the displacement described by $\boldsymbol{a}$.}
\label{fig:StrainCalc}
\end{figure}
Figure~\ref{fig:StrainCalc} illustrates the superposition of the displacements described by Eq.~\eqref{eq:A_vec_a}: If
we consider any mathematical point in the unstrained layer at the distance $x$ from the interface,
then this point has the distance $(1+\varepsilon_\mathrm{h})x$ from the interface after applying
$\mathds{1}\varepsilon_\mathrm{h}$. The displacement from this position into its final position is given by $\boldsymbol{a}x$.\\
If the layer is in static equilibrium, there is no stress perpendicular to the surface. Applying Hooke's law, this
constraint leads to the set of linear equations ($i=x,y,z$)
\begin{equation}
C_{ijkl}\varepsilon_{kl}n_j=\sum_{jkl}{\left(\varepsilon_\mathrm{h}\delta_{kl}+\frac{1}{2}(a_kn_l+a_ln_k)\right)C_{ijkl}n_{j}}=0,
\label{eq:EqSyst}
\end{equation}
where
\begin{equation}
\varepsilon_{kl}=\frac{A_{kl}+A_{lk}}{2}
\label{eq:Epsilon_sym}
\end{equation}
are the components of the (symmetrized) strain tensor and $C_{ijkl}$ denote the elastic stiffness constants of the crystal.
With given $C_{ijkl}$ and $\boldsymbol{n}$, Eq.~\eqref{eq:EqSyst} can be solved for the components $a_i$, which are proportional
to $\varepsilon_\mathrm{h}$. Considering Eq.~\eqref{eq:A_vec_a}, the distortion tensor $\bar{A}$ is then known as a linear
function of the only remaining parameter $\varepsilon_\mathrm{h}$, which can be determined experimentally as follows.\\
We label lattice planes by a single index $m$ which stands for a Miller index triplet, e.g.  $m=(hkl)$, and we sometimes
refer to planes by their normal $\boldsymbol{m}$. A lattice plane $m$ within a distorted layer in general encloses an angle
$\Delta \tau_m$ with the corresponding lattice plane in the substrate, cf. Fig.~\ref{fig:StrainCalc}. Furthermore, the distortion
causes a relative difference $(\Delta d/d)_m$ between the lattice plane spacings in layer and substrate. Both, $(\Delta d/d)_m$
and $\Delta \tau_m$, are accessible via HRXRD and are related to $\varepsilon_\mathrm{h}$ by the expressions\cite{JCG44_513}
\begin{eqnarray}
    \left(\frac{\Delta d}{d}\right)_m=\frac{d_m^\mathrm{l}\!-\!d_m^\mathrm{s}}{d_m^\mathrm{s}}=\frac{\boldsymbol{m}
    \cdot\boldsymbol{a}(\varepsilon_\mathrm{h})}{1+ \varepsilon_\mathrm{h}}
\cos\tau_m
\label{eq:Dd/d}\\   \Delta\tau_m=\frac{\boldsymbol{m} \cdot\boldsymbol{a}(\varepsilon_\mathrm{h})}{1+ \varepsilon_\mathrm{h}}
\label{eq:Dtau}\sin\tau_m.
\end{eqnarray}
\begin{figure}
\includegraphics[]{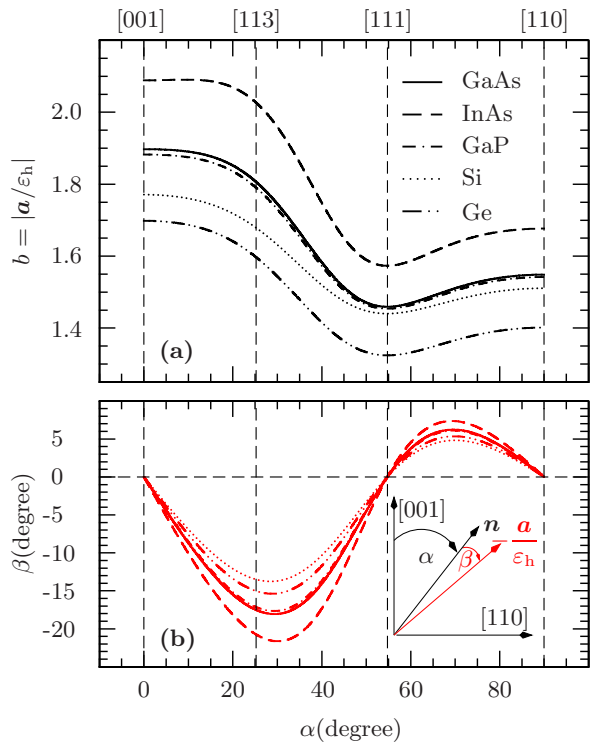}
\caption{(color online) (a) Magnitude $b=|\boldsymbol{a}/\varepsilon_\mathrm{h}|$ and (b) angle $\beta$, calculated with
various stiffness constants\cite{IOFFE}, are plotted as a function of the angle $\alpha$ between $\boldsymbol{n}$ and [001].
Only for the highly symmetric crystal orientations [001], [111], and [110] $\boldsymbol{a}$ aligns with $\boldsymbol{n}$.}
\label{fig:Vec_a}
\end{figure}
In the remainder of the paper, we focus on the practically relevant case of $(hhl)$-oriented substrates. Solving Eq.~\eqref{eq:EqSyst}
for substrate orientations $\boldsymbol{n}$ between [001] and [110], the vector $\boldsymbol{a}$ reads as
\begin{equation}
\boldsymbol{a}= - b\left(\begin{array}{c}
     \sin(\alpha+\beta)/\sqrt{2} \\
     \sin(\alpha+\beta)/\sqrt{2}\\
     \cos(\alpha+\beta)\\
     \end{array}\right)\varepsilon_\mathrm{h}\;,
\label{eq:vec_a_hhl}
\end{equation}
where $b$ denotes the magnitude of the vector $-\boldsymbol{a}/\varepsilon_\mathrm{h}$ and $\beta$ the angle between
$\boldsymbol{n}$ and $-\boldsymbol{a}/\varepsilon_\mathrm{h}$. In Fig.~\ref{fig:Vec_a}, $b$ and $\beta$ are plotted for several
cubic semiconductors as a function of the angle $\alpha$ between $\boldsymbol{n}$ and [001]. The vector $\boldsymbol{a}$ always
lies within the $(1\bar{1}0)$ plane; thus, this plain is a symmetry element (mirror plane) for all layers grown on $(hhl)$
substrates. Employing Eqs.~\eqref{eq:A_vec_a}, \eqref{eq:Epsilon_sym}, and \eqref{eq:vec_a_hhl}, the components of
$\bar{\varepsilon}$ can be inferred from Fig.~\ref{fig:Vec_a} using the equations
\begin{eqnarray}
\varepsilon_{xx}=\varepsilon_{yy}&=&\varepsilon_\mathrm{h}\left[1- \frac{b}{2}\sin(\alpha+\beta)\sin \alpha \right]\label{eq:eps_xx}\\
\varepsilon_{zz}&=&\varepsilon_\mathrm{h}\left[1- b\cos(\alpha+\beta)\cos \alpha \right] \label{eq:eps_yy}\\
\varepsilon_{xy}&=&\varepsilon_{xx}- \varepsilon_\mathrm{h}\label{eq:eps_xy}\\
\varepsilon_{xz}= \varepsilon_{yz}&=& -\frac{b\varepsilon_\mathrm{h}}{2\sqrt{2}}\left[\sin(\alpha+\beta)\cos \alpha \right.\nonumber \\
&+& \left. \cos(\alpha+\beta)\sin \alpha \right] \label{eq:eps_xz}.
\end{eqnarray}
Equations \eqref{eq:eps_xx}--\eqref{eq:eps_xz} are essential for the understanding of the MA and will be used in the derivation of the
free-energy density in Sec.~\ref{sec:MA}.\\
Figure \ref{fig:Vec_a} shows that for crystal facets other than (001), (110), and (111), $\boldsymbol{a}$ is not aligned with the
surface normal $\boldsymbol{n}$ and the layer is therefore sheared towards a direction given by the projection of $\boldsymbol{a}$
onto the surface (cf. Fig.~\ref{fig:StrainCalc}). Employing Eq.~\eqref{eq:Dtau}, the shear angle $\phi$ of a $\boldsymbol{n}$-oriented
layer towards any direction $\boldsymbol{p}\perp \boldsymbol{n}$, i.e. $\tau_p=90^\circ$, is obtained from
\begin{equation}
\phi=\Delta\tau_p=\frac{\boldsymbol{p}\cdot \boldsymbol{a}(\varepsilon_\mathrm{h})}{1+\varepsilon_\mathrm{h}},
\label{eq:Delta_t_p}
\end{equation}
where $\varepsilon_\mathrm{h}$ has to be determined by the procedure described above. $\phi$ can also be measured directly without making
use of the stiffness constants and the explicit form of $\boldsymbol{a}(\varepsilon_\mathrm{h})$, if we choose two lattice planes $m$
and $m'$ with $\boldsymbol{p}=(\boldsymbol{m}-\boldsymbol{m'})/2\sin\tau_m$; insertion into Eq.~\eqref{eq:Delta_t_p} yields
\begin{equation}
\phi =\frac{\boldsymbol{a}(\boldsymbol{m}-\boldsymbol{m'})}{2\sin\tau_m(1+ \varepsilon_\mathrm{h})}\underbrace{=}_{\mathrm{Eq.}~\eqref{eq:Dtau}}
\frac{\Delta\tau_m-\Delta\tau_{m'}}{2\sin^2\tau_m},
\label{eq:Phi}
\end{equation}
in agreement with Ref.~\onlinecite{JCG44_518}. The angles $\Delta \tau_m$ and $\Delta \tau_{m'}$ can be derived from rocking curves
as described in Sec.~\ref{subsec:HRXRD_exp}. Equation \eqref{eq:Phi} is valid if the two lattice planes $m$ and $m'$ are equally inclined
towards the surface, i.e. if $\tau_m=\tau_{m'}$, cf. Fig.~\ref{fig:StrainCalc}.\\
So far, we have restricted our considerations to pseudomorphically grown layers. With minor modifications, Eqs.~\eqref{eq:Dd/d} and
\eqref{eq:Dtau} can also be applied to partially relaxed layers. In Appendix \ref{Appendix_partially_relaxed}, we discuss how the
relaxed lattice constant and the degree of relaxation can be inferred from reciprocal space maps (RSMs) for arbitrarily oriented
substrates applying the formalism described above.

\subsubsection{Application to (113)-Oriented (Ga,Mn)As Layers\label{subsec:application}}
\begin{figure}[htp]
\includegraphics[]{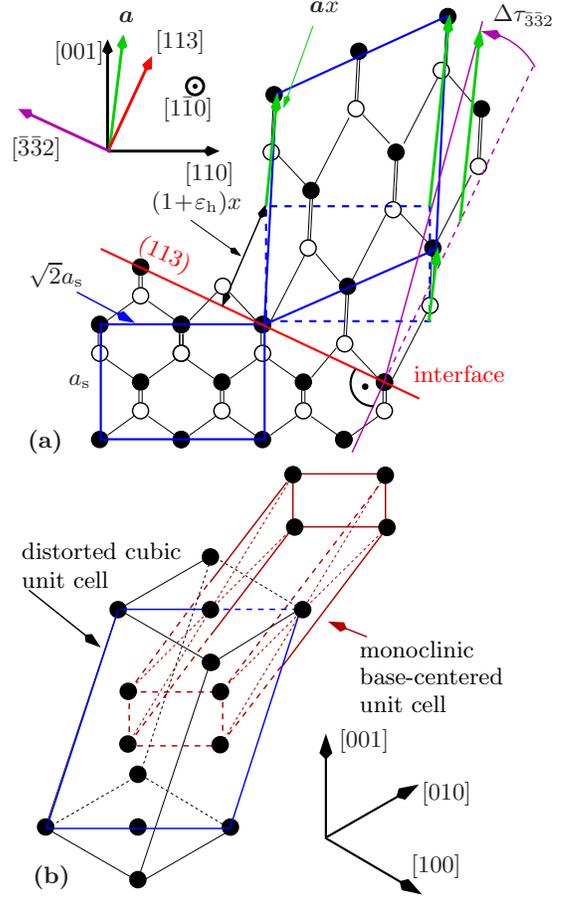}
\caption{(color online) (Ga,Mn)As(113)A layers exhibit a monoclinic crystal symmetry. (a) The layer as a whole is sheared towards
$[\bar{3}\bar{3}2]$ by an angle $\phi=\Delta\tau_{\bar{3} \bar{3}2}$, leaving the $(1\bar{1}0)$ mirror plane as the only remaining
symmetry element.\\
(b) 3-dimensional view of the monoclinic base-centered unit cell with respect to the distorted cubic unit cell. For the sake of
clarity, the second basis atom of the zinc-blende lattice is omitted in this sketch.}
\label{fig:Crystal_113}
\end{figure}

We now apply the general equations derived in the preceeding section to the case of (113)-oriented layers. For the following calculations,
we use the elastic stiffness constants of GaAs given in Ref.~\onlinecite{IOFFE} neglecting the Mn alloying. This assumption will be justified for (Ga,Mn)As layers
with Mn concentrations below 5\% by the experimental results presented in Sec.~\ref{subsec:HRXRD_exp}. From Fig.~\ref{fig:Vec_a} we obtain
for $\alpha=25.2^\circ$ the values $b=1.81$ and $\beta=-17.5^\circ$. Equation~\eqref{eq:vec_a_hhl} then yields
\begin{equation}
\boldsymbol{a}=-1.81\left(\begin{array}{c}
     0.095 \\
     0.095\\
     0.991 \\
     \end{array}\right)\varepsilon_\mathrm{h}.
    \label{eq:vec_a_113}
\end{equation}
With Eqs.~\eqref{eq:eps_xx}-\eqref{eq:eps_xz} we find for the strain tensor
\begin{equation}
    \bar{\varepsilon}=\varepsilon_{\mathrm{h}} ~ \left(\begin{array}{ccc}
     +0.95 & -0.05 & -0.35 \\
     -0.05 & +0.95 & -0.35\\
     -0.35 & -0.35 & -0.62 \\
     \end{array}\right).
    \label{eq:e_113}
\end{equation}
Figure \ref{fig:Crystal_113} schematically illustrates the crystal structure of the distorted (113) layer. The position of each atom in
the strained layer was constructed by applying Eq.~\eqref{eq:A_vec_a} together with Eq.~\eqref{eq:vec_a_113}: We start by putting a
hydrostatically compressed layer on the substrate. The unit cell of the layer in this hypothetical state is identical to the unit cell
of the substrate, cf. the dashed and solid cubic unit cells in Fig.~\ref{fig:Crystal_113} (a). The final position of each atom can be
found by a displacement along the direction given by $\boldsymbol{a}$, where the magnitude of the displacement is proportional to the
distance $x$ of the atom from the interface (cf. Fig.~\ref{fig:StrainCalc}).\\
The distortion breaks the symmetry of the layer and the only remaining symmetry element is a $(1\bar{1}0)$ mirror plane; hence, the
distorted crystal is assigned to the point group $m$ ($C_s$). This assignment is important for the derivation of the resistivity tensor
presented in Appendix \ref{Appendix_resistivity_tensors}. For clearness, the crystallographic unit cell of the distorted layer is
depicted in Fig.~\ref{fig:Crystal_113} (b); it becomes evident that the corresponding Bravais lattice is base-centered monoclinic.\\
In order to connect $\varepsilon_\mathrm{h}$ with the experimentally accessible quantity $(\Delta d/d)_m$, we apply Eq.~\eqref{eq:Dd/d}
to the case of the symmetric $m$=(113) reflection ($\tau_{113}=0^\circ$), i.e. we consider lattice planes parallel to the surface.
Equation \eqref{eq:Dd/d} simplifies to
\begin{equation}
\left(\frac{\Delta d}{d}\right)_{113}=-1.723 \frac{\varepsilon_\mathrm{h}}{1+\varepsilon_\mathrm{h}}=1.723 \frac{a_\mathrm{l}-a_\mathrm{s}}{a_\mathrm{s}}.
\label{eq:Dd/d_113}
\end{equation}
With $\varepsilon_\mathrm{h}$ as obtained from Eq.~\eqref{eq:Dd/d_113}, the shear angle of the (113) layer towards $[\bar{3}\bar{3}2]$ is inferred from Eq.~\eqref{eq:Delta_t_p}, reading now as
\begin{equation}
\phi=\Delta\tau_{\bar{3}\bar{3}2}=-0.544 \frac{\varepsilon_\mathrm{h}}{1+\varepsilon_\mathrm{h}}=0.544 \frac{a_\mathrm{l}-a_\mathrm{s}}{a_\mathrm{s}}.
\label{eq:Dt_m3m32}
\end{equation}
Figure~\ref{fig:Crystal_113} (a) tells us that this shear angle $\phi$ of the layer as a whole is the same as the angle
$\Delta\tau_{\bar{3}\bar{3}2}$ between the $(\bar{3}\bar{3}2)$ lattice planes of substrate and layer, respectively.\\

For a direct measurement of $\phi$, we use Eq.~\eqref{eq:Phi}.
\begin{figure}[htp]
\includegraphics[]{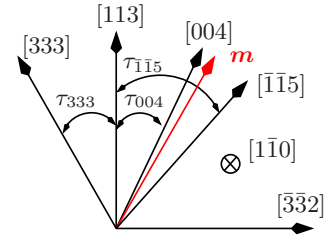}
\caption{(color online) Normal vectors of the lattice planes considered in Eq.~\eqref{eq:Phi_113}. Referring to Fig.~\ref{fig:StrainCalc},
we chose $\boldsymbol{m'}=[333]$ and $\boldsymbol{p}=[\bar{3}\bar{3}2]$. We find a virtual lattice plane $m$ with equal inclination with
respect to the (113) plane by interpolation between the $(004)$ and $(\bar{1}\bar{1}5)$ plane. For the corresponding angles we find
$\tau_m=\tau_{333}=\tau_{004}+0.27(\tau_{004}-\tau_{\bar{1}\bar{1}5})$.}
\label{fig:InterpolationAngle}
\end{figure}
Referring to Fig.~\ref{fig:InterpolationAngle} and Fig.~\ref{fig:StrainCalc}, we choose $m'=(333)$ and find the corresponding plane $m$
with equal inclination with respect to the (113) plane by interpolating between the $(\bar{1}\bar{1}5)$ plane and the (004) plane as
described in Ref.~\onlinecite{JCG44_518}. We obtain for the shear angle
\begin{equation}
\phi=\frac{\Delta\tau_{004}+0.27(\Delta\tau_{004}-\Delta\tau_{\bar{1}\bar{1}5})-\Delta\tau_{333}}{2\sin^2(\tau_{333})}\label{eq:Phi_113}.
\end{equation}

\subsection{Magnetic Anisotropy}\label{sec:MA}
MA is the dependence of a system's free-energy density $F$ on the orientation of the magnetization direction
$\boldsymbol{m}=\boldsymbol{M}/M$. In the following, we assume the sample to consist of a single ferromagnetic domain with a uniform
magnetization whose magnitude $M$ is assumed to be constant; we therefore analyze the quantity $F_M=F/M$.\\
For a phenomenological description of the MA in $(hhl)$-oriented (Ga,Mn)As layers, we expand $F_M$ in powers of the components $m_x$, $m_y$,
and $m_z$ of $\boldsymbol{m}$ along the cubic axes [100], [010], and [001], respectively. Considering terms up to the fourth order in
$\boldsymbol{m}$, the only intrinsic contribution to $F_M$ for an undistorted cubic layer is a cubic anisotropy
\begin{equation}
F_M^{\mathrm{cub.}}=B_{\mathrm{cub}}(m_x^4+m_y^4+m_z^4),
\label{eq:B_cubic}
\end{equation}
due to the crystal symmetry. Extrinsic contributions to $F_M$ are the shape anisotropy, caused by the demagnetization field perpendicular
to the layer, and a controversially discussed \cite{PRL90_167206,  PRB71_121302} uniaxial in-plane contribution
\begin{equation}
F_M^{\mathrm{ex.}}=B_\mathrm{d}(\boldsymbol{m}\cdot\boldsymbol{n})^2+B_{\mathrm{u}}(\boldsymbol{m}\cdot\boldsymbol{t})^2
\label{eq:B_extrinsic}
\end{equation}
where $\boldsymbol{n}$ and $\boldsymbol{t}$ denote unit vectors along the surface normal and $[\bar{1}10]$, respectively, and $B_\mathrm{d}$
is related to the magnetization $M$ by $B_\mathrm{d}=\mu_0M/2$.

For distorted layers, further intrinsic contributions proportional to the strain components $\varepsilon_{ij}$ occur. These are referred to
as magnetoelastic contributions and are presented for arbitrarily strained cubic crystals in Ref.~\onlinecite{VONSO_97}. Considering intrinsic
and extrinsic contributions, the free-energy density for $(hhl)$-oriented layers can be written as
\begin{eqnarray}
    F_M(\boldsymbol{m})&=& const. + B_{z^2}m_z^2+B_{xy}m_x m_y\nonumber\\&+&B_{xz}(m_x+m_y)m_z+B_{z^4}m_z^4
    +B_{x^4}(m_x^4+m_y^4)\nonumber\\&+&B_{xyz^2}m_x m_ym_z^2+B_{x^2yz}(m_x+m_y)m_xm_ym_z\nonumber\\
&+&B_\mathrm{d}(\boldsymbol{m}\cdot\boldsymbol{n})^2+B_{\mathrm{u}}(\boldsymbol{m}\cdot\boldsymbol{t})^2\label{eq:FE_113}.
\end{eqnarray}
The anisotropy parameters are related to the strain components by
\begin{eqnarray}
B_{z^2}&=&b_1(\varepsilon_{xx}- \varepsilon_{zz})\label{eq:B_z2}\\
B_{xy}&=&2b_2 \varepsilon_{xy}\label{eq:B_xy}\\
B_{xz}&=&2b_2 \varepsilon_{xz}\label{eq:B_xz}\\
B_{x^4}&=&b_4 \varepsilon_{xx}\!-\!\frac{3b_3+2b_4}{6}(2 \varepsilon_{xx}\!+\! \varepsilon_{zz})\!-\!\frac{B_\mathrm{cub}}{2}\label{eq:B_x4}\\
B_{z^4}&=&b_4 \varepsilon_{zz}\!-\!\frac{3b_3+2b_4}{6}(2 \varepsilon_{xx}\!+\! \varepsilon_{zz})\!-\!\frac{B_\mathrm{cub}}{2}\label{eq:B_z4}\\
B_{xyz^2}&=&2b_5 \varepsilon_{xy}\label{eq:B_xyz2}\\
B_{x^2z^2}&=&2b_5 \varepsilon_{xz}\label{eq:B_x2z2}.
\end{eqnarray}
The parameters $b_i$ denote magnetoelastic coupling constants. Using the trivial identity $m_x^2+m_y^2+m_z^2=1$, the contribution $B_{xy}m_xm_y$
in Eq.~\eqref{eq:FE_113} can be expressed in terms of $B_{\mathrm{u}}(\boldsymbol{m}\cdot\boldsymbol{t})^2$ and $B_{z^2} m_z^2$. Therefore, we will understand it to be contained in the latter terms in the analysis of the experimental data in Sec.~\ref{subsec:Exp}.

In the case of (001)-oriented layers, the off-diagonal elements of $\bar{\varepsilon}$ vanish and Eq.~\eqref{eq:FE_113} simplifies to the
well known expression
\begin{eqnarray}
    F_M(\boldsymbol{m})&=& const. + (B_{z^2}+B_\mathrm{d})m_z^2+B_{z^4}m_z^4\nonumber\\ &+&B_{x^4}(m_x^4+m_y^4)+
    \frac{1}{2}B_{\mathrm{u}}(m_x-m_y)^2\label{eq:FE_001}.
\end{eqnarray}

Equation \eqref{eq:FE_113} in particular explains in a natural way the occurrence of the uniaxial anisotropy  $B_{z^2}m_z^2$ along [001], which has been introduced
ad-hoc in previous publications in order to explain the results of angle-dependent ferromagnetic resonance\cite{PRB74_205205,APL89_012507} and
magnetotransport studies\cite{PRB74_205205,JAP103_093710} on (113)A-oriented (Ga,Mn)As layers.\\
Furthermore, Eqs.~\eqref{eq:B_z2}, \eqref{eq:B_x4}, and \eqref{eq:B_z4} account for the strain dependence of the parameters $B_{z^2}$, $B_{x^4}$,
and $B_{z^4}$\footnote{In Refs.~\onlinecite{PRB79_195206} and \onlinecite{PRB77_205210}, these parameters were labeled $B_{2\perp}$, $B_{4\parallel}$,
and $B_{4\perp}$, respectively.}, found in a systematic study of (Ga,Mn)As layers grown on relaxed (001)-oriented (In,Ga)As buffers\cite{PRB79_195206}: For (001)
orientation, the relation $\varepsilon_{xx}\approx-1.1\varepsilon_{zz}$ holds and therefore Eqs.~\eqref{eq:B_x4} and \eqref{eq:B_z4} read as
$B_{x^4}=(0.6b_3-0.7b_4)\varepsilon_{zz}-B_\mathrm{cub}/2$ and $B_{z^4}=(0.6b_3+1.4b_4)\varepsilon_{zz}-B_\mathrm{cub}/2$, respectively.
Figure 10 in Ref.~\onlinecite{PRB79_195206} shows an increase of  $B_{x^4}$ and a decrease of $B_{z^4}$ with increasing $\varepsilon_{zz}$. These
findings agree with Eqs.~\eqref{eq:B_x4} and \eqref{eq:B_z4} if $b_3>7b_4/6$ and $b_3,b_4<0$.

\subsection{Anisotropic Magnetoresistance}\label{sec:AMR}
It is well established that (similar to the MA) the AMR, described by the resistivity tensor $\bar{\rho}(\boldsymbol{m})$, is strongly affected by
the crystal symmetry. In order to obtain an analytical expression for $\bar{\rho}(\boldsymbol{m})$, we performed a symmetry-based series expansion
of the tensor components up to the fourth order in $\boldsymbol{m}$. For cubic and tetragonal symmetry, the explicit form of the resistivity tensor
and a detailed description of its derivation can be found in Ref.~\onlinecite{PRB77_205210}. In this work, we generalize the expressions for
$\bar{\rho}(\boldsymbol{m})$ to monoclinic and orthorhombic symmetry. As shown in Fig.~\ref{fig:orthorhombic} of Appendix \ref{Appendix_resistivity_tensors},
orthorhombic symmetry applies to (110)-oriented substrates. For other crystal facets $(hhl)$ with $0\neq h\neq l$, the crystal exhibits monoclinic
symmetry. The explicit forms of the corresponding tensors $\bar{\rho}(\boldsymbol{m})$ are given in Appendix~\ref{Appendix_resistivity_tensors}.

The AMR is usually probed by measuring the longitudinal and transverse resistivities $\rho_{\mathrm{long}}$ and $\rho_{\mathrm{trans}}$, respectively,
which are related to $\bar{\rho}(\boldsymbol{m})$ by $\rho_{\mathrm{long}}=\boldsymbol{j}^T\cdot\bar{\rho}\cdot \boldsymbol{j}$ and
$\rho_{\mathrm{trans}}=\boldsymbol{t}^T\cdot\bar{\rho}\cdot \boldsymbol{j}$. The unit vectors $\boldsymbol{j}$ and $\boldsymbol{t}$ point along the
current direction and the transverse direction, respectively. The resistivities also allow experimental access to the MA as shown in
Refs.~\onlinecite{PRB74_205205,PRB77_205210,PRB79_195206}. In Sec.~\ref{sec:MA}, the magnetic anisotropy parameters introduced in the preceding section
are derived experimentally by measuring the angular dependence of $\rho_{\mathrm{long}}$ and $\rho_{\mathrm{trans}}$ at various fixed magnetic field
strengths.

Now we turn to (113)-oriented layers. According to our previous work,\cite{PRB77_205210} we are referring to the right-handed coordinate system
$(\boldsymbol{j},\boldsymbol{t},\boldsymbol{n})$, where $\boldsymbol{j}\parallel [33\bar{2}]$, $\boldsymbol{t}\parallel [\bar{1}10]$, and
$\boldsymbol{n}\parallel[113]$; $m_j$, $m_t$, and $m_n$ denote projections of $\boldsymbol{m}$ along these directions. We calculate
$\rho_{\mathrm{long}}$ and $\rho_{\mathrm{trans}}$ by projecting the resistivity tensor in Eqs.~\eqref{eq:Resistivity_tensor_mono} along $\boldsymbol{j}$
and $\boldsymbol{t}$, respectively. We find
\begin{eqnarray}
\rho_{\mathrm{long}}&=&\rho_0+\rho_1m_j^2+\rho_2m_n^2+\rho_3m_j^4+\rho_4m_n^4+\rho_5m_j^2m_n^2\nonumber\\
                                            &+&\rho_{01}m_jm_n+\rho_{02}m_j^3m_n+\rho_{03}m_jm_n^3
\label{eq:RhoLong}
\end{eqnarray}
and
\begin{eqnarray}
    \rho_{\mathrm{trans}}&=& \rho_6m_n+\rho_7m_jm_t+\rho_8m_n^3\nonumber\\&+&\rho_9m_jm_tm_n^2+\rho_{10}m_j^3m_t+\rho_{11}m_tm_n\nonumber\\
                                            &+& \rho_{12}m_tm_n^3+\rho_{13}m_j+\rho_{14}m_j^3+\rho_{15}m_j^2m_n\nonumber\\
                                            &+& \rho_{16} m_jm_n^3 +\rho_{17}m_j^2m_tm_n, \label{eq:RhoTrans}
\end{eqnarray}
where $\rho_i$ are linearly independent resistivity parameters related to the expansion coefficients of $\bar{\rho}$.
In the limit of unstrained layers (cubic symmetry) the expressions for $\rho_\mathrm{long}$ and $\rho_\mathrm{trans}$ have the same form
as in Eqs.~\eqref{eq:RhoLong} and \eqref{eq:RhoTrans}, however several resistivity parameters become linearly dependent. This is due to the
fact that the current direction along $[33\bar{2}]$ already breaks the cubic symmetry.

AMR studies frequently focus on the special case where the magnetization lies within the layer plane ($m_n=0$). Eqs.~\eqref{eq:RhoLong} and
\eqref{eq:RhoTrans} then simplify to
\begin{eqnarray}    \rho_{\mathrm{long}}&=&\rho_0+\rho_1m_j^2+\rho_3m_j^4
\label{eq:RhoLong_ip}
\end{eqnarray}
and
\begin{eqnarray}
    \rho_{\mathrm{trans}}&=& \rho_7m_jm_t+\rho_{10}m_j^3m_t+\rho_{13}m_j+\rho_{14}m_j^3. \label{eq:RhoTrans_ip}
\end{eqnarray}

\section{Experiment\label{sec:experiment}}
We apply the theoretical expressions obtained in the preceding section to a series of (113)A(Ga,Mn)As layers with different Mn concentrations. In Sec.~\ref{subsec:HRXRD_exp}, the hydrostatic strain $\varepsilon_\mathrm{h}$ and thus $\bar{\varepsilon}$ as well as the shear angle of the layers are derived quantitatively from HRXRD measurements. In Sec.~\ref{subsec:HRXRD_exp}, we present angle-dependent magnetotransport measurements which are theoretically described by the expressions for $\rho_\mathrm{long}$ and $\rho_\mathrm{trans}$ given by Eqs.~\eqref{eq:RhoLong} and \eqref{eq:RhoTrans} and by the free-energy density given by Eq.~\eqref{eq:FE_113}.
\subsection{HRXRD Measurements\label{subsec:HRXRD_exp}}
\begin{figure}[htp]
\includegraphics[]{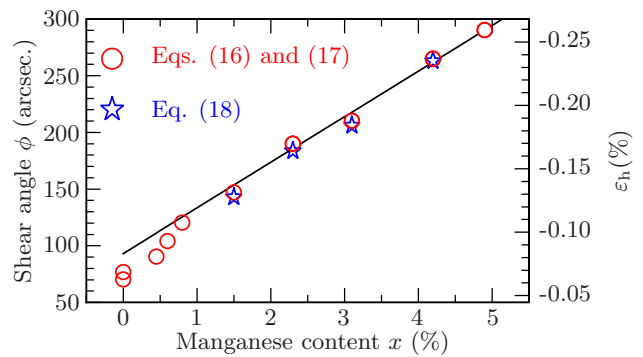}
\caption{(color online) Shear angle $\phi$ towards the $[\bar{3}\bar{3}2]$ direction and hydrostatic strain $\varepsilon_\mathrm{h}$ of (113)A-oriented
(Ga,Mn)As layers plotted as a function of the manganese concentration. The circles represent values for $\varepsilon_\mathrm{h}$ and $\phi$ derived
from $\omega$-$2\Theta$ scans using Eqs.~\eqref{eq:Dd/d_113} and \eqref{eq:Dt_m3m32}, respectively. The stars denote values for $\phi$ derived from
rocking curves applying Eq.~\eqref{eq:Phi_113}.}
\label{fig:ShearAngle}
\end{figure}

(Ga,Mn)As layers with manganese contents up to 5\% were grown on (113)A-oriented GaAs substrates by low-temperature molecular-beam epitaxy as
described in Refs.~\onlinecite{PRB71_205213} and \onlinecite{APL88_051904}. The structural properties of the (Ga,Mn)As layers were experimentally
investigated by HRXRD. We used a Bruker Siemens D5000HR x-ray diffractometer operating with the Cu-K$_{\alpha_1}$ radiation ($\lambda=0.154~\mathrm{nm}$).
In order to measure the strain $\varepsilon_\mathrm{h}$ and the thickness of the (Ga,Mn)As(113)A layers, we performed $\omega$-$2\Theta$ scans from the
symmetric (113) reflection. From the angular spacing of the layer thickness fringes\cite{JCG27_282}, we inferred a layer thickness of 150~nm.
Employing Bragg's law, the $\omega$-$2\Theta$ scan yields $(\Delta d/d)_{113}$, and via Eq.~\eqref{eq:Dd/d_113} we obtained the hydrostatic strain
$\varepsilon_\mathrm{h}$. With that value of $\varepsilon_\mathrm{h}$, we calculated the shear angle $\phi$ of the layer towards $[\bar{3}\bar{3}2]$
using Eq.~\eqref{eq:Dt_m3m32}. In Fig.~\ref{fig:ShearAngle}, $\varepsilon_\mathrm{h}$ and $\phi$ are plotted against the Mn content.\\
In order to verify the consistence of the formalism presented in Sec. \ref{subsec:distortion}, we measured the shear angle $\phi$ of several samples
directly by applying Eq.~\eqref{eq:Phi_113}. We inferred the angles $\Delta\tau_{333}$, $\Delta\tau_{004}$, and $\Delta\tau_{\bar{1}\bar{1}5}$ from
$\omega$-scans (rocking curves) with opened detector slits at high $(+)$ and low $(-)$ incidence. For an asymmetric reflection, as e.g. the (333)
reflection, the corresponding lattice plane encloses an angle $\tau_{333}$ with the surface, and the reflex can be measured at two different angles
$\omega_{333}^\pm=\Theta_{333} \pm \tau_{333}$ with respect to the surface, where $\Theta_{333}$ is the Bragg angle of the (333) reflection\cite{BARI_96}.
If the peak separation $\Delta \omega_{333}$ of layer and substrate is measured at high and low incidence, $\Delta \tau_{333}$ can be determined via
$\Delta\tau_{333}=(\Delta \omega_{333}^+-\Delta\omega_{333}^{-})/2$. In Fig.~\ref{fig:ShearAngle}, the results for the shear angle $\phi$ obtained
in this manner are shown in comparison to those derived from the $\omega$-$2\Theta$ scans. The excellent agreement of the values confirms the consistency
of the theoretical formalism. Furthermore, it demonstrates that the elastic stiffness constants of GaAs are a good approximation for those of (Ga,Mn)As
within the investigated range of Mn concentrations.

\subsection{Magnetotransport Measurements}\label{subsec:Exp}
Most of the samples under study were found to be insulating at $T=4.2$~K and could therefore not be investigated by magnetotransport.
The hole densities and Curie temperatures of the three conducting samples were determined from high-field magnetotransport measurements
as described in Ref.~\onlinecite{PRB80_125204}. The results are summarized in Table \ref{tab:Properties}.
\begin{table}[h]
\flushleft
\centering
\caption{Structural and electronic properties of metallic samples, studied via angle-dependent magnetotransport, cf. Sec.~\ref{sec:MA}.}
\begin{ruledtabular}
\begin{tabular}{c c c c c }
$x_\mathrm{Mn}(\%)$ & $\varepsilon_\mathrm{h}(\%)$ & $p(10^{20}\mathrm{cm}^{-3})$ & $T_\mathrm{C}$(K) \\\hline
4.9 &-0.26&2.0&44\\
4.2 &-0.23&2.2&47\\
3.1 &-0.18&1.5&38\\
\label{tab:Properties}
\end{tabular}
\end{ruledtabular}
\end{table}

\begin{figure}[htp]
\includegraphics[]{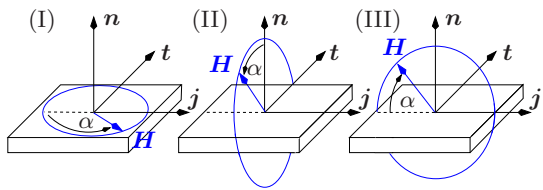}
\caption{(color online) Configuration (I), (II), and (III) for angle-dependent magnetotransport measurements. The external magnetic field
$\boldsymbol{H}$ was rotated within the planes normal to $\boldsymbol{n}\parallel[113]$, $\boldsymbol{j}\parallel[33\bar{2}]$, and
$\boldsymbol{t}\parallel[\bar{1}10]$.}
\label{fig:Config}
\end{figure}

\begin{figure*}[htp]
\includegraphics[]{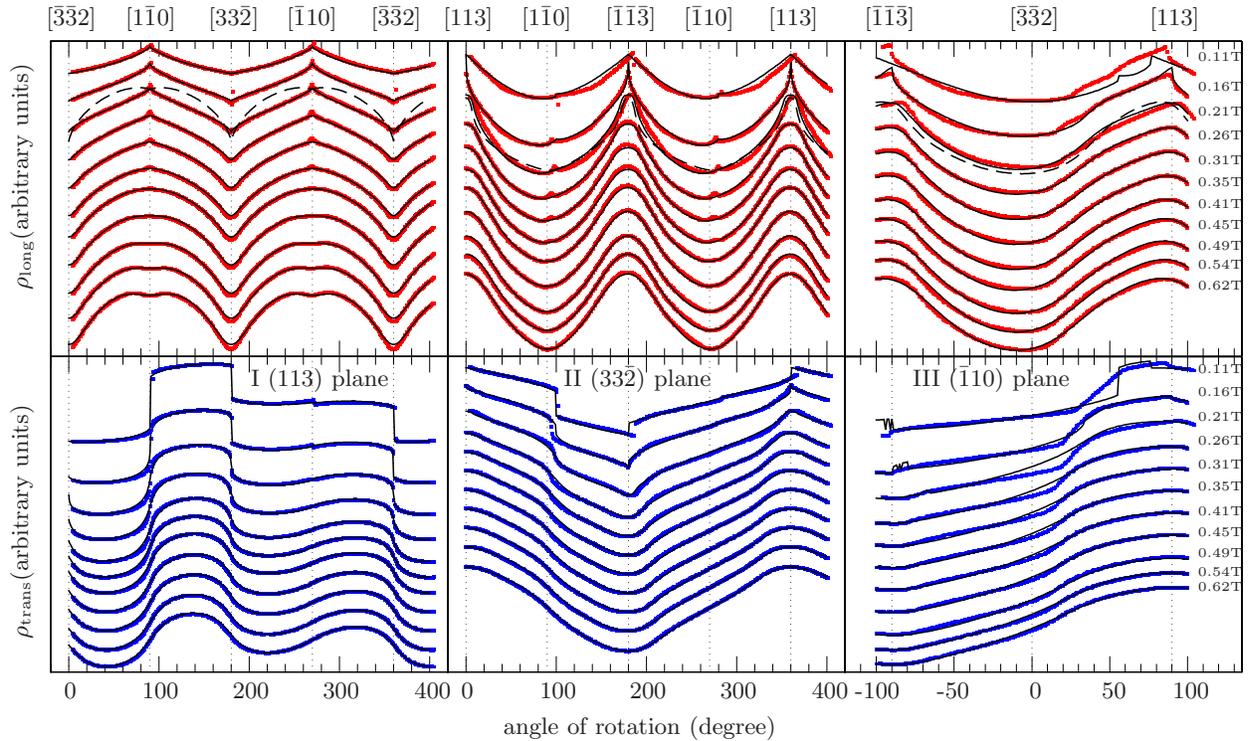}
\caption{Angular dependence of the longitudinal and transverse resistivity at various external magnetic field strengths rotated within three
different planes, referred to as configuration (I), (II), and (III), respectively (cf. Fig.~\ref{fig:Config}). The sample contains nominally
4.9\% Mn and the hydrostatical strain is $\varepsilon_\mathrm{h}=-0.26\%$. The longitudinal-resistivity curves can be simulated
using $H$-dependent resistivity parameters, cf. Fig.~\ref{fig:B516_H}. In particular, for configuration (I) the longitudinal-resistivity parameters
strongly vary with the field strength $H$. For comparison, simulated curves with the unchanged resistivity parameters as obtained at $\mu_0H=0.62~\mathrm{T}$ are shown for $\mu_0H=0.21~\mathrm{T}$ (dashed lines). The transverse-resistivity parameters are found to be constant within the experimental magnetic
field range.}
\label{fig:B516_res}
\end{figure*}

In order to investigate the MA and AMR in the (113)A-oriented (Ga,Mn)As samples, we performed angle-dependent magnetotransport
measurements.\cite{PRB74_205205,PRB77_205210,PRB79_195206} For this purpose, the samples were patterned into 0.3~mm-wide Hall-bar structures
oriented along $[33\bar{2}]$ with Ohmic Au-Pt-Ti contacts and the longitudinal voltage probes separated by 1~mm. The dc-current density was
$220~\mathrm{Acm}^{-2}$. The samples were mounted on a rotatable sample holder in a liquid-He-bath cryostat, which was placed between the poles
of a LakeShore electromagnet. With this setup, the magnetic field could be rotated arbitrarily with respect to the crystallographic axes of the
(Ga,Mn)As layer.\\
We rotated the external field $\boldsymbol{H}$ at various fixed field strengths within the three different crystallographic planes depicted in
Fig.~\ref{fig:Config} and measured $\rho_\mathrm{long}$ and $\rho_\mathrm{trans}$.
In the presence of an external magnetic field $\boldsymbol{H}$, the free-enthalpy density $G_M=G/M$ instead of the
free-energy density $F_M$ determines the magnetization orientation. We thus write
\begin{equation}
G_M(\boldsymbol{H})=F_M(\boldsymbol{m})-\mu_0\boldsymbol{m}\cdot \boldsymbol{H},
\label{eq:G_113}
\end{equation}
with $F_M(\boldsymbol{m})$ from Eq.~\eqref{eq:FE_113}. The orientation $\boldsymbol{m}$ of the magnetization at a given external field
$\boldsymbol{H}$ can be found by minimizing Eq.~\eqref{eq:G_113} with respect to $\boldsymbol{m}$. At the maximum applied field of
$\mu_0H= 0.62~\mathrm{T}$, the Zeeman term $\mu_0\boldsymbol{H}\boldsymbol{m}$ dominates the free-enthalpy density and $\boldsymbol{m}$ essentially
aligns along $\boldsymbol{H}$. With decreasing field strength however, the MA described by the anisotropy parameters in Eq.~\eqref{eq:FE_113}
more and more governs the motion of the magnetization as $\boldsymbol{H}$ is rotated with respect to the sample.\\
By fitting Eqs.~\eqref{eq:RhoLong} and \eqref{eq:RhoTrans} to our experimental data recorded at $\mu_0H=0.62~\mathrm{T}$, we obtained values
for the resistivity parameters $\rho_i$. Using these parameters, we simulated the measured angular dependencies of $\rho_\mathrm{long}$ and
$\rho_\mathrm{trans}$ at weaker fields by varying the anisotropy parameters until the simulated curves fit the experiment.
Figure~\ref{fig:B516_res} exemplarily shows the experimental and simulated angular dependencies of $\rho_\mathrm{long}$ and $\rho_\mathrm{trans}$
for the sample with 4.9\% Mn.\\

The experimentally obtained anisotropy parameters are listed in Table~\ref{tab:AI_parameters}. In agreement with other
work,\cite{JAP103_093710,PRB74_205205} the MA in our samples can be described by the parameters $B_{x^4}$, $B_{z^4}$,  $B_{z^2}$, $B_{\mathrm{d}}$,
and $B_\mathrm{u}$. As expected, the shape anisotropy parameter $B_\mathrm{d}=\mu_0M/2$ increases with increasing Mn concentration. Since both,
the Mn concentration and the epitaxial strain, influence the parameters $B_{x^4}$, $B_{z^4}$, and $B_{z^2}$, it is not possible to infer an
unambiguous strain-dependence of these parameters from our experiment. Nevertheless, some qualitative conclusions can be drawn:\\
Because the relations $B_{x^4}-B_{z^4}=b_4~(\varepsilon_{xx}-\varepsilon_{zz})\approx 1.57~ b_4~\varepsilon_\mathrm{h}>0$ and
$\varepsilon_\mathrm{h}<0$ hold for all samples, we find a negative magnetoelastic coupling parameter $b_4$ in agreement with the discussion
in Sec.~\ref{sec:MA}. Since the parameters $B_{xz}$, $B_{x^2z^2}$, and $B_{xyz^2}$ are negligible (the influence of $B_{xy}$ is contained in $B_{z^2}$ and $B_\mathrm{u}$ due to the linear
dependence mentioned earlier), we are led to the conclusion that the coupling parameters $b_2$ and $b_5$ are smaller than $b_1$ (at least for
the samples with $x_\mathrm{Mn}>4 \%$). For more strongly strained samples or for samples with larger coupling parameters $b_i$, however, all
anisotropy parameters given in Eqs.~\eqref{eq:B_z2}--\eqref{eq:B_xyz2} may play a role.\\
\begin{table}[htp]
\flushleft
\caption{Anisotropy parameters of the samples under study as obtained from angle-dependent magnetotransport measurements.}
\begin{ruledtabular}
\begin{tabular}{c c c c c c c }
$x_\mathrm{Mn}(\%)$ & $B_{x^4}(\mathrm{mT})$ & $B_{z^4}(\mathrm{mT})$ & $B_{z^2}(\mathrm{mT})$ & $B_{\mathrm{d}}(\mathrm{mT})$ &
$B_\mathrm{u}(\mathrm{mT})$\\\hline
4.9 &-46&-11&20&18&-8 \\
4.2 &-56&-25&20&12&-8\\
3.1 &-65&-40&5&5&-5\\
\label{tab:AI_parameters}
\end{tabular}
\end{ruledtabular}
\end{table}
\begin{figure}[htp]
\includegraphics[]{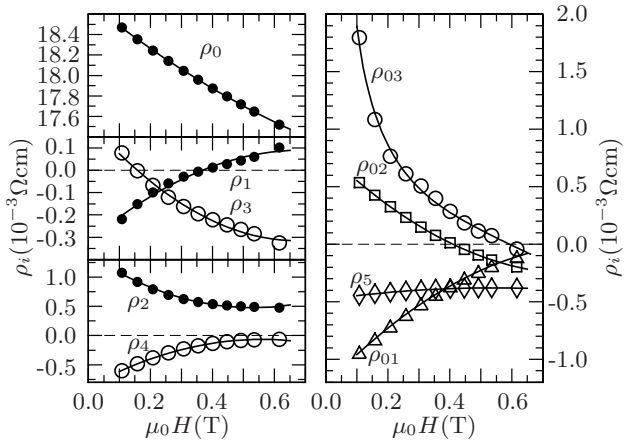}
\caption{Dependence of the longitudinal resistivity parameters $\rho_0$ -- $\rho_{03}$ on the field strength $\mu_0H$ for a (Ga,Mn)As(113)A
layer with a Manganese concentration of 4.9\%. The decrease of $\rho_0$ with increasing $\mu_0H$ reflects the negative magnetoresistance. The
lines are guides to the eye.}
\label{fig:B516_H}
\end{figure}
In contrast to our previous experiments,\cite{PRB74_205205, PRB77_205210} the drastic change of the longitudinal-resistivity curves, in particular
those obtained in configuration (I), upon variation of the external field strength (cf. Fig.~\ref{fig:Config} and \ref{fig:B516_res}) cannot solely be
explained by MA. A satisfactory agreement between theory and experiment can only be obtained by allowing for field-dependent longitudinal-resistivity
parameters. In Fig.~\ref{fig:B516_H}, the best-fit longitudinal-resistivity parameters of the sample with 4.9\% Mn are plotted as a function of the magnetic
field. The longitudinal-resistivity parameters of the samples with 4.2\% and 3.1\% Mn showed a similar field dependence. In contrast, the transverse
resistivities shown in Fig.~\ref{fig:B516_res} can be simulated with field-independent resistivity parameters $\rho_{6}-\rho_{17}$. In order to obtain a good fit of the experimental data, all parameters in Eq.~\eqref{eq:RhoTrans} with exception of $\rho_9$ and $\rho_{17}$ are required. The variation of
the lineshapes upon the field strength exclusively arises from the MA described by the anisotropy parameters from Eq.~\eqref{eq:FE_113}. Therefore, we
mainly focused on the transverse resistivities when deriving MA parameters. Field-dependent resistivity parameters have also been reported by other
groups for (113)A-oriented (Ga,Mn)As\cite{JAP103_093710} and for (001)-oriented (Ga,Mn)As\cite{PRB77_125320}. In Ref.~\onlinecite{PRB77_125320} the
field dependence was studied up to 9~T.\\
The microscopic origin of these findings is not clear yet. A (001)-oriented reference sample, grown at the same conditions as the (113)A-oriented
layer with $x_\mathrm{Mn}=4.9\%$, showed a similar field dependence of the longitudinal resistivities, indicating that the effect is not primarily
related to the substrate orientation.
\section{Summary}\label{summary}

Starting from a continuum mechanical treatment of the lattice distortion in high-index epilayers, a general expression for the strain
tensor $\bar{\varepsilon}$ of $(hhl)$-oriented layers was derived. The isotropic strain component $\varepsilon_\mathrm{h}$ (and thus
$\bar{\varepsilon}$) as well as the shear angle $\phi$ were related to the experimentally accessible quantities $\Delta d/d$ and $\Delta\tau$.
Applying the equations to the special case of (113)A orientation, $\varepsilon_\mathrm{h}$ and $\phi$ could be experimentally determined
for a series of (113)A-oriented (Ga,Mn)As layers using HRXRD. Based on symmetry considerations, analytical expressions for the free-energy
density and the resistivity tensor were derived by means of series expansions in terms of the magnetization components up to the fourth order, allowing
for a phenomenological description of the MA and AMR, respectively. The anisotropy parameters were explicitly given as a function of the
strain-tensor components. The expression for the resistivity tensor, deduced for monoclinic and orthorhombic crystal symmetry, can be used to
calculate the longitudinal and transverse resistivities for arbitrary current directions. In order to probe the MA and AMR of the (Ga,Mn)As
samples by angle-dependent magnetotransport, expressions for the resistivities were derived for current direction along [33$\bar{2}$].
The measurements were performed at 4.2 K and revealed the presence of a strong uniaxial anisotropy $B_{z^2}m_{z^2}$ along [001] which could
be explained within our theoretical model by the explicit form of $\bar{\varepsilon}$. Further significant contributions to the MA were found
to be $B_{z^4}m_z^4$, $B_{x^4}(m_x^4+m_y^4)$, $B_\mathrm{d}(\boldsymbol{m}\cdot\boldsymbol{n})^2,$ and
$B_\mathrm{u}(\boldsymbol{m}\cdot\boldsymbol{t})^2$. Whereas the transverse resistivity parameters turned out to be nearly constant within the
range of applied magnetic fields, the longitudinal resistivity parameters were found to strongly depend on the field strength.

\appendix
\section{Partially Relaxed Layers\label{Appendix_partially_relaxed}}
\begin{figure}[htp]
\includegraphics[]{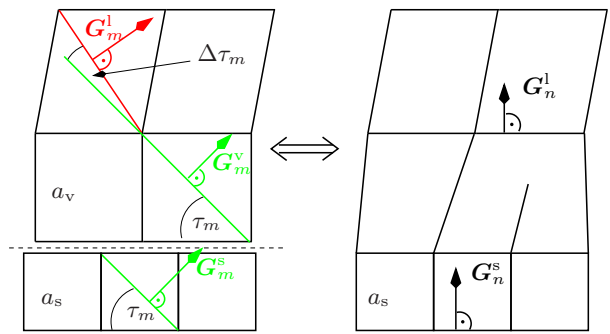}%
\caption{(color online) A partially relaxed layer can be thought of as a layer that pseudomorphically grows on a virtual cubic substrate with a lattice constant $a_\mathrm{v}$ different from the true substrate lattice constant $a_\mathrm{s}$. Lattice planes $m$ are parallel in real and virtual substrate. The relative inclination $\Delta \tau_m$ of the plane $m$ in the layer with respect to $m$ in the virtual substrate is the same as the inclination with respect to $m$ in the real substrate.\\ If the layer was tilted with respect to the substrate the reciprocal lattice vectors $\boldsymbol{G}_n^\mathrm{s}$ and $\boldsymbol{G}_n^\mathrm{l}$ of a symmetric reflection would not be parallel. Note that in this schematic the cubes do not necessarily represent the cubic unit cells.}
\label{fig:virtualSubstrate}%
\end{figure}

In this Appendix, we describe how the equations presented in Sec. \ref{sec:structure} can be used to characterize partially relaxed layers. Figure \ref{fig:virtualSubstrate} illustrates that the layer in the partially relaxed state can be described as a layer which is commensurate with a virtual cubic substrate having a lattice constant $a_\mathrm{v}$.
The hydrostatic strain in the layer is then described by the parameter
\begin{equation}
\varepsilon_\mathrm{h}^*=\frac{a_\mathrm{v}-a_\mathrm{l}}{a_\mathrm{l}},
\label{eq:epsilon_h_star}
\end{equation}
where $a_\mathrm{l}$ is the relaxed lattice parameter of the layer.
In order to describe partially relaxed layers, we have to replace $\varepsilon_\mathrm{h}$ by $\varepsilon_\mathrm{h}^*$ in all equations of Sec. \ref{sec:structure}.
In particular, we obtain
\begin{eqnarray}
    \left(\frac{\Delta d}{d}\right)_m=\frac{d_m^\mathrm{l}\!-\!d_m^\mathrm{v}}{d_m^\mathrm{v}}=\frac{\boldsymbol{m} \cdot\boldsymbol{a}(\varepsilon_\mathrm{h}^*)}{1+ \varepsilon_\mathrm{h}^*}
\cos\tau_m
\label{eq:Dd/d_star}\\  \Delta\tau_m=\frac{\boldsymbol{m} \cdot\boldsymbol{a}(\varepsilon_\mathrm{h}^*)}{1+ \varepsilon_\mathrm{h}^*}
\label{eq:Dtau_star}\sin\tau_m.
\end{eqnarray}
instead of Eqs.~\eqref{eq:Dd/d} and \eqref{eq:Dtau}.

From a reciprocal space map (RSM) around an asymmetric reflex ($m \neq n$), $\Delta\tau_m$ can be inferred, because it is the angle between the reciprocal lattice vector of layer $\boldsymbol{G}_m^\mathrm{l}$ and substrate $\boldsymbol{G}_m^\mathrm{s}$, respectively, cf. Fig.~\ref{fig:virtualSubstrate}. $\varepsilon_\mathrm{h}^*$ can then be obtained from Eq.~\eqref{eq:Dtau_star}. Assuming that the lattice parameter of the (real) substrate is known, the reciprocal lattice of the substrate can serve as a reference; this allows an accurate measurement relative to the substrate without relying on the absolute angle scale of the diffractometer.
The length of the layer's reciprocal lattice vector $\left|\boldsymbol{G}_m^\mathrm{l}\right|$ can be inferred from the RSM and consequently the lattice plane spacing of the layer $d^\mathrm{l}_m=2\pi/ \left|\boldsymbol{G}_m^\mathrm{l}\right|$
is obtained. By inserting $\varepsilon_\mathrm{h}^*$ into Eq.~\eqref{eq:Dd/d_star}, we find the value of $d^\mathrm{v}_m$. Because the virtual substrate is cubic, we obtain $a_\mathrm{v}=d^\mathrm{v}_m \sqrt{h^2+k^2+l^2}$ and with Eq.~\eqref{eq:epsilon_h_star} we find $a_\mathrm{l}$. Thus we can determine the degree of relaxation
\begin{equation}
R=\frac{a_\mathrm{v}-a_\mathrm{s}}{a_\mathrm{l}-a_\mathrm{s}}.
\label{eq:Degree_of_relaxation}
\end{equation}

If the relaxed layer is tilted with respect to the substrate, as it has been reported for relaxed (In,Ga)As layers grown on (001)GaAs \cite{PRB79_195206}, this tilt needs to be considered in the determination of $\varepsilon_\mathrm{h}^*$. The tilt angle $\psi = \angle (\boldsymbol{G}_n^\mathrm{l},\boldsymbol{G}_n^\mathrm{s})$ can be inferred from a RSM around a symmetric reflection ($m=n$), and the corrected angle $\Delta\tau_m^\mathrm{corr}=\Delta\tau_m-\psi$ has to be inserted into Eq.~\eqref{eq:Dtau} in order to obtain $\varepsilon_\mathrm{h}^*$.

\section{Resistivity tensor for monoclinic and orthorhombic symmetry\label{Appendix_resistivity_tensors}}
\begin{figure}[htp]
\includegraphics[]{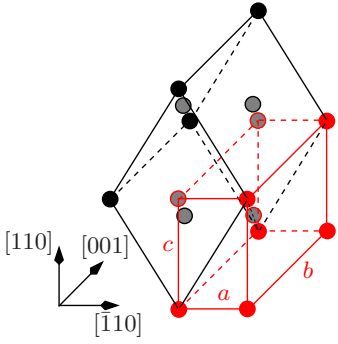}%
\caption{(color online) Distortion of the fcc cubic unit cell (black lines) when a layer with relaxed lattice constant $a_\mathrm{l}$ is grown on a (110) plane of a substrate with lattice constant $a_\mathrm{s}<a_\mathrm{l}$.  The grey points denote the face centered atoms. For the zinkblende lattice the only symmetry elements are a twofold rotational axis around [001] and two $\sigma_v$ planes; thus the point group is $mm2$ ($C_{2v}$). The crystallographic unit cell of the body centered orthorhombic Bravais lattice is shown in red. Assuming pseudomorphic growth, the relations between the orthorhombic lattice parameters and the cubic lattice constant of the substrate are $a=a_\mathrm{s}\sqrt{2}/2$ and $b=a_\mathrm{s}$.}
\label{fig:orthorhombic}%
\end{figure}

We derived the resistivity tensors for monoclinic and orthorhombic symmetry up to the fourth order in $\boldsymbol{m}$; thereby we made use of von Neumann's principle as described in Ref. \onlinecite{PRB77_205210}. In Fig.~\ref{fig:orthorhombic}, we show that for (110)-oriented substrates the layer exhibits orthorhombic symmetry.
For this case the generating matrices are
\begin{equation}
    \bar{S}_3=\left(\begin{array}{ccc}
     -1 & 0 & 0 \\
     0 & -1 & 0\\
     0 & 0 & 1\\
     \end{array}\right).
    \label{eq:S_3}
\end{equation}
and
\begin{equation}
    \bar{S}_5=\left(\begin{array}{ccc}
     0 & 1 & 0 \\
     1 & 0 & 0\\
     0 & 0 & 1\\
     \end{array}\right).
    \label{eq:S_5}
\end{equation}
Note that here the matrix $\bar{S}_5$ has been adapted to the cubic frame of reference, where the symmetry operation is a reflection at the $xy$ plane and not a reflection at the $y$ plane as in the canonical representation for the matrix $\bar{S}_5$.
We obtain for the resistivity tensor
\begin{equation}
\bar{\rho}_\mathrm{orthorhombic}=\bar{\rho}_\mathrm{tetragonal}+\Delta \bar{\rho}_\mathrm{orthorhombic},
\label{eq:Resistivity_tensor_ortho}
\end{equation}
where $\bar{\rho}_\mathrm{tetragonal}$ is given by Eqs.~(3), (4), and (5) in Ref. \onlinecite{PRB77_205210} and $\Delta \bar{\rho}_\mathrm{orthorhombic}$ by Eq.~\eqref{eq:Delta_rho_ortho}, respectively.\\
For monoclinic symmetry, the only generating matrix is $\bar{S}_5$ and we find
\begin{equation}
\bar{\rho}_\mathrm{monoclinic}=\bar{\rho}_\mathrm{orthorhombic}+\Delta \bar{\rho}_\mathrm{monoclinic},
\label{eq:Resistivity_tensor_mono}
\end{equation}
where $\Delta\bar{\rho}_\mathrm{monoclinic}$ is given by Eq.~\eqref{eq:Delta_rho_mono}. The Greek letters in Eq.~\eqref{eq:Delta_rho_ortho} and \eqref{eq:Delta_rho_mono} are non-vanishing linear combinations of the galvanomagnetic tensors, cf. Eq.~(2) in Ref. \onlinecite{PRB77_205210}.

\begin{widetext}
\begin{eqnarray}
\Delta \bar{\rho}_\mathrm{orthorhombic} & = &
    \left(\begin{array}{ccc}
            0        &    \alpha_1 &    0 \\
            \alpha_1 &    0        &    0 \\
            0        &    0        &    0
    \end{array}\right)
    +
    \left(\begin{array}{ccc}
            0             &    0            &    \beta_1 m_x  \\
            0             &    0            &    -\beta_1 m_y \\
            -\beta_1 m_x  &    \beta_1 m_y  &    0
    \end{array}\right) \nonumber\\
    & + &
    \left(\begin{array}{ccc}
            0                      & \gamma_1 (m_x^2+m_y^2) &    0 \\
            \gamma_1 (m_x^2+m_y^2) & 0                      & 0 \\
            0                      & 0                      &    0
    \end{array}\right)
    +
    \left(\begin{array}{ccc}
            \gamma_2 m_x m_y        &    0                   & \gamma_4 m_y m_z    \\
            0                       &    \gamma_2 m_x m_y    & \gamma_4 m_x m_z    \\
            \gamma_4 m_y m_z     &    \gamma_4 m_x m_z & \gamma_3 m_x m_y
    \end{array}\right) \nonumber\\
    & + &
    \left(\begin{array}{ccc}
            0                 &    0                 &    \delta_1 m_x^3  \\
            0                 &    0                 &    -\delta_1 m_y^3 \\
            -\delta_1 m_x^3   &    \delta_1 m_y^3    &    0
    \end{array}\right)
    +
    \left(\begin{array}{ccc}
            0                     & \delta_2 m_x m_y m_z &    0 \\
            -\delta_2 m_x m_y m_z & 0                    &    0    \\
            0                     & 0                    &    0
    \end{array}\right) \nonumber\\
    & + &
    \left(\begin{array}{ccc}
            0 &
            0 &
            \delta_{3} m_x m_y^2       \\
            0&
            0&
            -\delta_{3} m_x^2 m_y  \\
            -\delta_{3} m_x m_y^2   &
            \delta_{3} m_x^2 m_y    &
            0
    \end{array}\right) \nonumber\\
    & + &
    \left(\begin{array}{ccc}
            0                           &    \epsilon_1 (m_x^4 + m_y^4) &    0\\
            \epsilon_1 (m_x^4 + m_y^4)  &    0                          &    0\\
            0                           &    0                          &    0
    \end{array}\right)
    +
    \left(\begin{array}{ccc}
            0                         &    \epsilon_2 m_x^2 m_y^2 &    0\\
            \epsilon_2 m_x^2 m_y^2 &    0                         &    0 \\
            0                         &    0                         &    0
    \end{array}\right) \nonumber\\
    & + &
    \left(\begin{array}{ccc}
            \epsilon_3 m_x^3 m_y  + \epsilon_4 m_x m_y^3 &
            0&
            \epsilon_5 m_y^3 m_z                          \\
            0 &
            \epsilon_3 m_x m_y^3  + \epsilon_4 m_x^3 m_y   &
            \epsilon_5 m_x^3 m_z                          \\
            \epsilon_5 m_y^3 m_z &
            \epsilon_5 m_x^3 m_z &
            \epsilon_6 m_x m_y (m_x^2 + m_y^2)
    \end{array}\right) \nonumber\\
    &+&
    \left(\begin{array}{ccc}
            0                           &    0                           &    \epsilon_7 m_x^2 m_y m_z \\
            0                           &    0                           &    \epsilon_7 m_x m_y^2 m_z \\
            \epsilon_7 m_x^2 m_y m_z &    \epsilon_7 m_x m_y^2 m_z &    0
    \end{array}\right)
    \label{eq:Delta_rho_ortho}
\end{eqnarray}
\begin{eqnarray}\Delta \bar{\rho}_\mathrm{monoclinic}&=&
 \left(
    \begin{array}{ccc}
        0       &   0 & \alpha_2\cr
        0 & 0        &  \alpha_2\cr
        \alpha_2 &  \alpha_2  & 0
    \end{array}
\right ) +
 \left( \begin{array}{ccc}
        0                          &    \beta_2(m_x+m_y)          & \beta_3 m_z                             \cr
        -\beta_2 (m_x+m_y)         &    0                         & - \beta_3 m_z             \cr
         - \beta_3 m_z             &     \beta_3 m_z              & 0
\end{array} \right ) \nonumber\\
& + &
\left( \begin{array}{ccc}
        0                               &   0                               &   \gamma_5 m_x^2 + \gamma_6 m_y^2\cr
        0                               &   0                               &   \gamma_5 m_y^2 + \gamma_6 m_x^2\cr
        \gamma_5 m_x^2 + \gamma_6 m_y^2 & \gamma_5 m_y^2 + \gamma_6 m_x^2   &   0
\end{array} \right ) \nonumber\\
& + &
\left(
    \begin{array}{ccc}
        \gamma_8 m_x m_z + \gamma_9 m_y m_z                   & \gamma_{10} m_z (m_x + m_y)                              & \gamma_7 m_x m_y                             \cr
        \gamma_{10} m_z (m_x+m_y)                                &  \gamma_8 m_y m_z + \gamma_9 m_x m_z                   & \gamma_7 m_x m_y                             \cr
        \gamma_7 m_x m_y                                      & \gamma_7 m_x m_y                                      & \gamma_{11} m_z (m_x+m_y)
    \end{array}
\right ) \nonumber\\
& + &
\left(
    \begin{array}{ccc}
        0                         & \delta_4 (m_x^3 + m_y^3) &  0               \cr
        -\delta_4 (m_x^3 + m_y^3) & 0                        &  0               \cr
        0                         & 0                        &  0
    \end{array}
\right )
+
\left(  \begin{array}{ccc}
        0                     & 0                    &  \delta_5 m_x m_y m_z \cr
        0                     & 0                    &  -\delta_5 m_x m_y m_z\cr
        -\delta_5 m_x m_y m_z & \delta_5 m_x m_y m_z &  0
    \end{array}
\right )\nonumber \\
& + &
\left(  \begin{array}{ccc}
        0 &
        \delta_8 (m_x^2 m_y + m_x m_y^2)&
        \delta_6 m_x^2 m_z + \delta_7 m_y^2 m_z      \cr
        -\delta_8 (m_x^2 m_y + m_x m_y^2)&
        0&
        - \delta_6 m_y^2 m_z - \delta_7 m_x^2 m_z  \cr
        - \delta_6 m_x^2 m_z - \delta_7 m_y^2 m_z &
        \delta_6 m_y^2 m_z + \delta_7 m_x^2 m_z&
        0
    \end{array}
\right )\nonumber \\
& + &
\left(  \begin{array}{ccc}
        0&  0                     & \epsilon_8 m_x^4 + \epsilon_9 m_y^4\cr
        0                         & 0&  \epsilon_8 m_y^4 + \epsilon_9 m_x^4\cr
        \epsilon_8 m_x^4 + \epsilon_9 m_y^4&    \epsilon_8 m_y^4 + \epsilon_9 m_x^4&    0
    \end{array}
\right )
+
\left(  \begin{array}{ccc}
        0&  0 & \epsilon_{10} m_x^2 m_y^2\cr
      0  &  0&  \epsilon_{10} m_x^2 m_y^2 \cr
        \epsilon_{10} m_x^2 m_y^2 & \epsilon_{10} m_x^2 m_y^2 & 0
    \end{array}
\right ) \nonumber\\
& + &
\left(
    \begin{array}{ccc}
        \epsilon_{11} m_x^3 m_z + \epsilon_{12} m_y^3 m_z &
        \epsilon_{15} m_z (m_x^3 + m_y^3)&
        \epsilon_{13} m_x^3 m_y  +\epsilon_{14} m_x m_y^3                           \cr
        \epsilon_{15} m_z (m_x^3 + m_y^3) &
        \epsilon_{11} m_y^3 m_z + \epsilon_{12} m_x^3 m_z &
        \epsilon_{13} m_x m_y^3  + \epsilon_{14} m_x^3 m_y                           \cr
        \epsilon_{13} m_x^3 m_y  + \epsilon_{14} m_x m_y^3  &
        \epsilon_{13} m_x m_y^3  + \epsilon_{14} m_x^3 m_y  &
        \epsilon_{16} m_z (m_x^3 + m_y^3)
    \end{array}
\right ) \nonumber\\
&+&
\left(  \begin{array}{ccc}
        \epsilon_{17} m_x^2 m_y m_z + \epsilon_{18} m_x m_y^2 m_z & \epsilon_{19} m_x m_y m_z (m_x + m_y) & 0 \cr
        \epsilon_{19} m_x m_y m_z (m_x + m_y) & \epsilon_{17} m_x m_y^2 m_z + \epsilon_{18} m_x^2 m_y m_z & 0\cr
        0                         &  0 &    \epsilon_{20} m_x m_y m_z (m_x + m_y) \cr
    \end{array}
\right )
\label{eq:Delta_rho_mono}
\end{eqnarray}
\end{widetext}

\begin{acknowledgments}
This work was supported by the Deutsche Forschungsgemeinschaft under contract
number Li 988/4.
\end{acknowledgments}

\end{document}